# A model for urban growth processes with continuum state cellular automata and related differential equations

**Alberto Vancheri, Paolo Giordano, Denise Andrey, Sergio Albeverio[1]**

Accademia di architettura, via Canavée, CH- 6850 Mendrisio
avancheri@arch.unisi.ch

## Abstract

A new kind of cellular automaton (CA) for the study of the dynamics of urban systems is proposed. The state of a cell is not described using a finite set, but by means of continuum variables. A population sector is included, taking into account migration processes from and towards the external world. The transport network is considered through an integration index describing the capability of the network to interconnect the different parts of the city. The time evolution is given by Poisson distributed stochastic jumps affecting the dynamical variables, with intensities depending on the configuration of the system in a suitable set of neighbourhoods. The intensities of the Poisson processes are given in term of a set of potentials evaluated applying fuzzy logic to a practical method frequently used in Switzerland to evaluate the attractiveness of a terrain for different land uses and the related rents. The use of a continuum state space enables one to write a system of differential equations for the time evolution of the CA and thus to study the system from a dynamical systems theory perspective. This makes it possible, in particular, to look systematically for bifurcations and phase transitions in CA based models of urban systems.

## 1. Introduction

### State of the art

There are many good reasons for which urban systems can be considered, at least for some of their features, as complex dynamical systems. Self organization, pattern formation, emergency of global properties from local interactions are among the most typical aspects that urban systems share with a wide class of complex systems studied in mathematics, natural and social sciences (Batty and Longley, 1994; O'Sullivan and Torrens, 2000; Torrens, 2001). Cellular automata (CA) are among the most powerful and flexible computational tools developed in order to capture some features of complex systems (Margolus, 1987 and references therein). One of the most appealing features of CA in modelling urban systems is their intrinsic spatial structure. A real city is indeed characterized by a spatial distribution of different land uses competing and cooperating through interactions depending on the spatial proximity (Langlois and Phipps, 1997; Portugali, 1999; Semboloni, 2000; White and Engelen, 1997; White et al. 2000). Furthermore the dynamics of CA is not directed by a 'headquarter' where the decision over the evolution of the system are taken, but is rather the result of cooperative or antagonist behaviour among the different parts of the system at the local level. These are not the only aspects relevant for urban systems though they play a very important role (other aspects being provided by demographical and economical global factors or by the obvious influence of a centralized planning of the city (Langlois and Phipps, 1997; Portugali, 1999; White and Engelen, 1997).

One of the most useful and powerful CA for the study of urban growth processes has been developed by R. White and G. Engelen (1997, 2000).
In the last few years many problems have been raised that require, however, new modelling strategies. The first one concerns the calibration of the many parameters appearing in the most developed and powerful CA models, as for example that of White and Engelen (Bäck et al., 1996). The problem is connected with the possibility to compare simulated configurations of the city with the empirical ones, and thus it is also connected with the important problem of models' validation (i.e. how can we answer precisely the question:

---

[1] Universität Bonn, Institut für Angewandte Mathematik, Abt. Stochastik, Wegelerstr. 6, D-53115 Bonn, BiBoS (Bielefeld-Bonn-Stochastic Research Centre), IZKS (International Centre for the Study of Complex Systems).

"in what sense does this stochastic model faithfully describe the real system?"). This is part of a more general problem, connected with the fact that the state of a CA is often only a *qualitative* object, not suitable for powerful *analytical* investigation methods. Qualitative or categorical maps can be effectively compared using fuzzy logic methods (Power et al., 2001; Hagen, 2003). But the use of numerical variables instead of categorical maps enables one to apply fuzzy logic to a set of suitable chosen quantitative indicators to base the comparison of maps (see Vancheri et al. b, in preparation). Another problem is connected with the expected presence of bifurcations and phase transitions in complex systems. These are highly typical phenomena of complex systems that have been extensively studied in the field of dynamical systems, especially those described by differential equations (see Weidlich, 2000; Guckenheimer and Holmes, 1997; Kuznetsov, 1998; Torrens, 2001; Wilson, 1981; Haag 1986, Holyst et al., 2000 and references therein).

The dependence of the observed time evolution of the urban system on the initial configuration and the response of the system to an external perturbation are other examples of problems that are hard to be faced in an effective way in the conceptual framework of CA. In fact such problems are mainly discussed in the framework of dynamical systems described by differential equations. To try to associate an ordinary differential equation to a CA one must at least have a quantitative description of the configuration of the CA through *numerical variables*. Furthermore one must be able to take a limit for the discrete time step going to zero. Exogenous effect are often also discussed in relation with the theory of control of continuous dynamical systems.

A large amount of work in this direction has been done in the last 30 years in the field of synergetics and socio-dynamics (see Haken, 1978; Haag, 1989; Schweitzer, 2002a and 2002b; Weidlich and Haag, 1983; Haag and Weidlich, 1984; Weidlich, 1991, 1997 and 2002 and references therein). In this approach a socio economical system is described through a master equation, governing the time evolution of the probability distribution on the configuration space of the system. Under certain conditions a differential equation for the time evolution of the mean values of the dynamical variables can be derived. Bifurcations, chaotic dynamics, phase transitions and other interesting behaviours are effectively observed in many of these models of socio-economical systems [ibidem and e.g. Munz and Weidlich, 1990a and 1990b; Weidlich and Haag, 1987; Popkov, Shvetsov and Weidlich, 1998].

## Features of our model

In this paper we propose a new model for the study of urban growth processes with the aim of providing a suitable conceptual and computational frame taking into account above considerations.
The main features of the model are:

i) The state of a cell is not given by a *qualitative* label chosen in a finite set of possibilities, as in the more usual CA models, but by a *continuum* set of real numbers. Let us recall that some continuum state CA have been already employed in the study of complex systems (see e.g. Ostrov, 1997).

ii) The model contains a set of stochastic *evolution rules* containing explicitly the duration $\Delta t > 0$ of the time step. This results in a stochastic process based on stochastic jumps of the state variables.

iii) There are *several kinds of neighbourhoods* describing different kinds of urban interactions at different space scales.

iv) Each cell is uniquely determined by a label $c \in \Gamma$, where $\Gamma \subset \mathbb{Z}^2$ is the cellular space, and *may have an "arbitrary shape"*. In some urban models it is natural to choose the shape of the cells taking into account the subdivision of the urban space determined by urban plans or by some feature of the land associated with the structure of the settlement or with the topography of the land (e.g railway, motorways, rivers, lakes).

v) In case of a Markovian dynamics (that is in case we are considering urban transformation without memory), in the limit for $\Delta t \to 0$ we derive a system of *ordinary differential equations* for the mean values of the dynamical variables. More generally, even in case of a non-Markovian dynamics, we derive instead a system of random differential equations, which uses the concept of forward mean derivative (Nelson, 1985), for the extensive state variables.

The main aim of the work is to join the powerful approach of CA models, in particular its simple and detailed description of the urban system and of the forces acting in it, with the advantages given by the mathematical and conceptual tools developed in the field of synergetics and socio-dynamics, first of all the possibility to describe the system using differential (or random differential) equations.

This permits a multi-level approach in the construction of the model: on the one hand we have a deep foundational background on mathematical and physical theories, and on the other hand we have a detailed and meaningful construction of the model from the point of view of urban studies. The model is indeed constructed as an attempt to formalize the real behaviour of the urban (stochastic) processes rather than to capture some abstract aspects in the dynamics of the city through an over idealized model.

Moreover the continuum state space permits to obtain *quantitative* outputs from computer simulations. Hence the problem of calibration of the model can also be faced in a more effective way: because the configuration of the system is given by a set of numerical variable in a Euclidean space $\mathbb{R}^d$, we can use some familiar distance concept to compare configurations, e.g. the Euclidean one or the percentage of deviation with respect to the final configuration of the real urban system.

In other words we have both the usual simplicity of the CA approach in the construction of a model, through the definition of simple but meaningful *local* rules, and the advantages of having differential equations which we derive for the dynamic of the state variables.

Our approach enables us to provide a mathematical framework where it is easier to face basic problems in urban planning. E.g. we can use the model to study the behaviour of the urban system after the building of an important infrastructural work, supporting in this way the decision process. The model we propose makes a large use of tools concretely employed in urban planning, as for example the set of indexes that constrain the exploitation of the spatial resources of the city, and so it suggests answers to practical as well as to theoretical problems.

From the practical point of view one can handle the question of the effectiveness of a given urban planning strategy. We can use the model e.g. to decide whether the control of the urban quality by given indexes leads or not to a real improvement of the quality of life. In our model we can look for planning strategies which do not work against the forces acting in the urban system. From the theoretical point of view this is connected to the theory of optimal control (Davis, 2002; Hernandez-Lerma and Lasserre, 1999; Elliot et al., 1992; Hernandez-Lerma, 1989; Kushner, 1971; Lee and Markus, 1967) that is to the interplay between self organization and centralized control of a complex system. Our model is suited to incorporate such methods in natural way.

## Structure of the article

We will present the model starting from a "generic" urban region. In this way we aim at describing the characteristics that the data of a city must have and how they have to be aggregated, managed and simplified to prepare the input data of the model. For simplicity we will say that these input data describe an "idealized" city. In this way we want also clarify that we are doing some assumptions which enable a simplification in the model. These simplified assumptions could however be improved without changing deeply the general structure of the model. In subsequent works (Vancheri et al. b, in preparation) we will use our model to perform computer simulations on real case studies and we will expose some strategies concerning calibration and validation of the model.

We shall distinguish general modelling strategies and definitions from particular choices, so that we shall describe the model by a *decreasing* level of generality: e.g. we will say "to define a stochastic evolution rule of the CA one has to define a probability density of this type…" and only subsequently "in our case we choose the following probability density…". We hope in this way to clarify where one can replace, for some particular purpose, our assumptions by different ones.

In section 2 we present the idealized urban system describing its land uses (2.1), its urban plan (2.2) and our formalization of the transportation network (2.3).

In section 3 we define the cellular automaton (CA): we start with the cellular decomposition of the space, the system of neighbourhoods and the borders (3.1); next we give the state space of a cell (3.2), and thus the dynamical variables, and finally we define the evolution rules which correspond to the urban processes we are considering (3.3).

In section 3.4 we introduce the probability distributions associated to our CA, in particular we explain the urban assumptions which conduct us to assume that the number of events that happen in a cell is a Poisson process. A basic idea of our model is that the intensities ('number of events per time unit') of these Poisson processes depend on the state of the neighbourhoods: the more the state of the neighbourhoods is in favour of these events, the higher are the related intensities. In this section we also introduce the probability distributions of the stochastic goods produced in elementary events.

In section 4 we introduce the general approach for the construction of the intensities. In 4.1 we show how we face the problem of the interplay between the local, neighbourhoods depending effects and the global rate of growth of the different land uses. Even if we distinguish in some cases these two levels of the dynamics, we avoid to treat them separately, fixing exogenously (or quasi exogenously) the global rates of growth. In 4.2 we introduce the *potential,* a basic concept that has proved to be useful in order to describe in the language of our model the forces acting in the urban space. These quantities are interpreted as truth values of statements (expressed in natural language) describing the local attractiveness of the space for a given elementary process. We have used fuzzy logic, fuzzy set theory and fuzzy decision making methods in order to connect the potentials with the configuration of the neighbourhoods of cells.

In section 5 we give a description of the evolution algorithm of the CA Finally in section 6 we introduce the system of ordinary differential equations associated to the model using an intuitive, but clear and meaningful, deduction. For rigorous proofs concerning this section and the derivation of the random differential equations mentioned above we refer to Vancheri et al. a (in preparation) and Giordano et al. (in preparation).

## 2. From real data to input data

In this section we will introduce a description of the data we need in order to apply our model: what are the different types of land uses we consider, the types of buildings, what land uses do not evolve endogenously in the model, the indexes used to define the urban plan's laws and the formalization of the transport network into the model. We refer to an imaginary (but realistic) urban system that we have employed for the computer simulations that will be presented in Vancheri et al. b (in preparation).

### 2.1. Land uses

We consider a region consisting of a medium-sized city surrounded by some smaller towns, agriculture and free terrain at disposal for future expansions of the urban system.
We consider the following set of land uses: residences ($R$), commerce ($C$), offices ($O$), industry ($I$), agriculture ($A_g$) and secondary urbanisation structures ($U$). Beside these land uses there are other fixed typologies of terrain that do not evolve endogenously with time, like rivers, woods, railway, streets. We distinguish three kinds of buildings:

1) residential buildings, containing spaces that can be used indifferently for offices or residences. In some cases the first floor of these buildings is occupied by small commercial surfaces as shops, boutiques, restaurants, and so on that we will indicate with $CI$ (commerce of kind $I$).
2) Commercial buildings, constituted by big commercial surfaces located in shop centres that we will indicate with $CII$ (commerce of kind $II$).
3) Factories.

Of course a part of the urbanized region can be occupied by infrastructures as green parks, parking places, schools, sport centres, theatres, administration offices. We shall not model endogenously the dynamics of these land uses; furthermore we will aggregate all these infrastructures in just one category called *secondary urbanisation works*. This aggregation is certainly a rough approximation. Nevertheless, the distinction among different kinds of secondary urbanisation structures does not require a change in the modelling strategy but just the introduction of more terms in the functions describing the dynamics of the processes we consider.

## 2.2. The urban plan

Our model is conceived in a way that makes it easy to implement the constraints given by urban plans. In the present model we are particularly inspired by Swiss legislation, but the model can be easily modified for different situations.

The urban plan we are considering subdivides the region in small areas (for the sake of simplicity they will be identified with the cells used for the CA) where two quantities are defined: the *edification index* $I_M^e$ and the *covering index* $I_M^c$ defined in the following way:

$$I_M^e := \frac{V^M}{S_T}, \qquad I_M^c := \frac{S^M}{S_T} \qquad (2.2.1)$$

where $S_T > 0$ denotes the difference between the total surface of the cell and the surface occupied by fixed elements as rivers, railways and streets; $S^M$ is the maximal surface that can be covered by buildings in the cell and $V^M$ the maximal volume that can be built in the cell. These indexes, giving constraints to the exploitation of the spatial resources in a cell, are used by the administration to regulate the quality of the urban space and the features of the urbanisation process, which can be more or less extensive. The index '$M$' means "maximal". Definitions (2.2.1) can be interpreted as an upper bound on the exploitation of spatial resources. If in a cell a volume $V$ has already been built on a covered surface $S$, there remains an amount $I_M^e \cdot S_T - V$ and $I_M^c \cdot S_T - S$ respectively of volume and surface resource for further development.

## 2.3. The communication network

The *connections* of the region with the external world are given by motorways, railways, rivers or airports. These principal connections are distinguished in the model from the *local transportation network* which can consist in a regional network of streets or in a regional railways system. We stress this difference because in the first case (connection with the external world) the effect of the connection on the urban system can be described by a function giving the accessibility of each cell to the access point of the network (e.g. railways stations, motorways exits) measured through distance or accessibility times. The regional system, being much more articulated in its structure, requires a more precise method in order to evaluate the capability of the different access point to integrate the different parts of the urban space. For this end we will describe the network as a graph and we will use the *integration index* of a vertex in order to evaluate the effectiveness of the vertex in integrating the urban space (this quantity is defined in the theory of graphs and used in configurational analysis in a rather different way (Hillier and Hanson, 1984). We start modelling the network as a graph, constituted by a finite set of vertices $V$ connected by edges $E$. The vertices of the graph are the access point to the network. The edges are the connection lines. Given two vertices $v, w \in V$ we define the distance $d(v,w)$ as the minimum number of edges needed to move from $v$ to $w$. This distance must be distinguished from the length of the shortest path between $v$ and $w$, because it is defined in a topological way counting edges. In the case of a subway, for example, this distance is interpreted as the number of railway stops needed to move from a station to another. Inserting fictitious vertices in the graph in such a way that we find a vertex every $l$ meters along each edge, we can consider the topological distance $d$ as an approximation of the metric distance up to a length $l$. This latter method can be useful also when a sort of spatial impedance, due for example to traffic, is associated to the lines of the network. In this case the vertices can separate stretches characterized by a similar value of the spatial impedance. Using the additivity of the spatial impedance we can interpret the distance defined above as a measure of the spatial impedance separating two points of the network.

The *integration index* of a vertex $v \in V$ is often described in the following way:

$$i'(v) := \frac{\sum_{v' \in V} d(v,v')}{N-1} \qquad (2.3.1)$$

where $N$ is the number of vertices in the graph. In order to avoid size effects we will use the following normalized expression for the integration index in place of the previous one (Steadman, 1983):

$$i(v) := \frac{2 \cdot (i'(v) - 1)}{N - 2} \quad (2.3.2)$$

It is easy to show that the normalized index takes values between 0 and 1. Vertices with a low value of the integration index are characterized by a short average distance from all the other vertices of the graph. Thus the relevance of this index consists in distinguishing the most integrated vertices from the most segregated ones. We will assume that the proximity with a well integrated vertex is an appealing feature for some land uses as residences and commerce. The method used to associate an integration value to cells rather than to vertices is explained briefly in 4.2, where fuzzy logic techniques are introduced.

The use of the integration index can be criticized observing that it describe only to what extent the network interconnects the parts of the system, without taking into account what is connected. Of course the use of this index is based on the assumption that there is a certain coherence in the way the urban system and the transport network have grown together in the past. We must assume for example that to well integrated vertices there correspond, at least as a general tendency, important parts of the city. This assumption has to be tested through a preliminary "field" study where for example the location of some important structures is compared on a map with the integration indexes associated to the transport network.

## 3. The structure of the cellular automaton

In order to define the structure of the cellular automaton we have to introduce the following elements:

i) a cellular decomposition of the space
ii) the time step and the system of neighbourhoods associated to cells
iii) the type of borders
iv) a configuration space to describe the state of the cells
v) a system of rules to describe the change of the state of the cells from one step to the next
vi) the probability distributions of the stochastic processes described by the CA
vii) the algorithm which connects these probabilities with the previous rules.

We will face this task in the next sections.

### 3.1. Cellular decomposition, time step neighbourhoods and borders

The first step in constructing our model of an urban system consists in a cellular decomposition of the space. We will identify each *cell* (here with this term we mean a delimited given portion of the urban area) with two indexes $c = (i, j) \in Z \subseteq \mathbb{Z}^2$ and we will simply say that $c = (i, j)$ is a *cell* of the automaton. We will always suppose $Z$ given by a finite subset of $\mathbb{Z}^2$.

As in usual CA models the time is discrete. As mentioned above, the quantities relevant for the dynamics of the system contain explicitly the duration $\Delta t > 0$ of the temporal step as a parameter. In this way we can study the behaviour of the model with respect to different concrete choices of $\Delta t$ and we can study the "speed of changing" of these quantities as $\Delta t$ goes to 0, obtaining in the limit a differential equation. We will see later (see section 3.3) that explicit criteria for the choice of the time step $\Delta t$ are available (about this problem see also Bäck et al., 1996).

The implementation of evolution rules, as in usual CA models, is based upon the specifications of a neighbourhood associated to each cell $c$. We prefer to use a system of neighbourhoods $U^j(c) \subseteq Z$, $j = 1, \cdots, n$ instead of only one.

Working with a system of neighbourhoods is just a way to take into account the fact that different interactions among land uses take place at different spatial scales and are mediated by different kind of connections. The effect of distance is described in our model by weights associated to each neighbourhood and depending on the kind of spatial interaction which is considered. If for example we are considering the disturbance due to the noise, a function of the usual metric distance seems to be a relevant to measure the effect of distance. But if we are considering the access to a given service from a residential site, a more complicated object must be considered, taking into account for example the effectiveness of the connection with respect to the kind of service. The subway for example can be a good kind of connection to reach the place where one works, but not to reach a shopping centre; in this latter case a connection by car might be more suitable. We will give more details about the definition of the system of neighbourhoods in section 4.2 where the above mentioned weights are more clearly connected to the definition of another important quantity: the potential.

The situation in which it is most simple to handle borders for our type of CA is the one given by an urban region with natural fixed borders, like rivers, woods, motorways, or surrounded by a large, free terrain available for urban growth; here "large" means "with respect to the maximum growth we are able to forecast in the time interval we are considering. In this case one has to take the cellular lattice $Z$ sufficiently large and to assign a fixed value to the state of the border. Of course it becomes important to verify that the dynamics of the CA is not essentially influenced by the particularly chosen values, e.g. checking that there is a negligible statistical correlation between the state of the border and the state of every cell. Situations where the border cannot be considered as fixed, e.g. when we consider a part of a city embedded in a larger urban system, are not easy to manage. In these cases the system interacts with the external world not only through global interactions as the ones connected with the demographical pressure, but also precisely "through its border". In these cases more complicated models have to be developed in order to describe how the local dynamics of the space outside the border affects the evolution of the system.

### 3.2. State space of a cell

– **Dynamical variables**

The state of a cell in usual CA models for urban studies is given by a qualitative label, describing the prevalent land use in the cell. As stated in the introduction we prefer to use a set of numbers giving information about *how* different land uses occupy the space inside the cell. In our model we have chosen the following quantities to define the state of a cell:

- The amount of *built* volume for residences and offices (indicated with the common label $A$), for small commercial surfaces ($CI$) for big commercial surfaces ($CII$) and for industries ($I$). We will indicate these quantities respectively by $V^A$, $V^{CI}$, $V^{CII}$ and $V^I$.
- The volumes $V_u^R, V_u^{CI}, V_u^O$ *effectively in use* respectively for residences, commerce $CI$ and offices (the difference $V^X - V_u^X$ being the unsold or unused volume of the use $X$).
- The surfaces $S^{Ag}$ used for agriculture.
- The surface $S$ covered by buildings (we do not distinguish here the different kinds of buildings).
- The surface resource $\Delta$ already employed and hence no more available for further development.
- The surface $\Delta^A$ used for residential buildings (including private gardens, parking, and infrastructure).

All these quantities are expressed by real numbers (relative to chosen units of surface and volume).
Let us give some remarks about the dynamical variables:

i) We do not distinguish here built volume for residences and offices because we suppose that these land uses can be located in the same kind of space.
ii) We will assume that the commercial surfaces of type $I$ are first built and then sold or given for rent. They can be occupied and abandoned following a dynamic depending on the level of the demand and rents. Commercial surfaces of kind $II$ are built and then directly used by big commercial companies and, as a consequence, follow different dynamics than those of type I. Due to the fact that we do not consider unused commercial surfaces of kind II, we do not specify among the variables the amount of commercial surfaces of kind II in use. For the same reason we do not consider here unused industrial spaces.

iii) The quantity $S_F = S_T - \Delta$, where $S_T$ is the total surface of the cell computed without considering the fixed areas (e.g. water, railways), gives the amount of surface at disposal in the cell for further development.

iv) Empirical data about the urban region have to be chosen having in mind these dynamical variables. We remark that some of them can be estimated using aerial views of the region.

Grouping together the eleven dynamical quantities listed above, we obtain the state vector $v(c) \in \mathbb{R}^d$, (in this case $d = 11$) associated to the cell $c$:

$$v(c) = \left(V^A, V^{CI}, V^{CII}, V^I, V_u^R, V_u^{CI}, V_u^O, S^{Ag}, S, \Delta^A, \Delta\right) \quad (3.2.1)$$

When we need to explicitly indicate the time dependence of the state vector we will use the notation $v(c,t)$. We call $E_c = \mathbb{R}^d$ the state space of a cell of the CA. A vector in this space substitutes the discrete label used in more conventional CA models. The state space of the whole system is obtained summing up the vector spaces of all the cells:

$$E = \mathbb{R}^D = \oplus_{c \in \Gamma} E_c \quad (3.2.2)$$

where $D = n \cdot d$ and $n$ is the number of cells in $\Gamma$.

– **Local parameters**

The description of a cell needs to be completed specifying a set of quantities that do not have an autonomous dynamics. Some of them can vary exogenously. We will call them *local parameters* and we will group them in a second vector $w(c)$ associated to the cell $c$.

The most important quantities belonging to $w(c)$ and considered in the model are the following:

i) the integration index $i(v)$ of the vertex $v$ of the local transport network which is nearest to the cell $c$ (see section 2.3),
ii) the distance of the cell from the nearest vertex of the local transport network,
iii) the distances from the railway, the motorway, the motorway exits and the river to take into account several effects as the disturbance due to the noise and the availability of infrastructure for the transportation of the goods.
iv) The edification and the covering indexes given by the urban plan (see section 2.2),
v) the surface occupied by secondary urbanizations works,
vi) the surface occupied by fixed elements as rivers, railways, streets and other infrastructures.

Depending on what urban quantities we are interested in, we can put in the vector $w(c)$ besides these principal local parameters also topographical information as the slope of the terrain.

The use of the vectors $v(c)$ and $w(c)$ discards all the information about the spatial organization of the land uses which lies under the scale of the cell. Nevertheless the state vector contains very detailed information about the urban typology of the cell. For example one can reconstruct the mean height $\overline{h}(c)$ of the buildings in a cell $c$ through the simple formula:

$$\bar{h}(c) = \frac{V^A(c) + V^{CII}(c) + V^I(c)}{S(c)} \qquad (3.2.3)$$

(actually this is a weighted mean with weights given by the fraction of surface occupied by A, CII, and I respectively). A further example is the local density of the settlement measured through the actual value of the edification index:

$$I^V(c) = \frac{V^A(c) + V^{CII}(c) + V^I(c)}{S_T(c)} \qquad (3.2.4)$$

or the actual value of the covering index

$$I^S(c) := \frac{S(c)}{S_T(c)} \qquad (3.2.5)$$

As a further example let us consider the component $V_u^R$ of the state vector. This variable is connected with the populations sector of the model. It gives in fact an estimation of the population resident in the cell through the occupied residential volume. Comparing the built volume $V^R$ with the effectively used one $V_u^R$ we obtain some information about the dynamics of the construction sector. Indeed we will assume for example that a large amount of unused residential volume can discourage the construction of new buildings.

The subdivision of the relevant quantities for the descriptions of the cells into dynamical variables and local parameters is related to the idea underlying the use of this kind of model. Some of the local parameters associated to a cell can be modified acting on the infrastructural system of the city, some others through changes in the urban plans. *In all these cases we do not model the dynamics of the related processes, but assume them as exogenous.* This is equivalent to consider the parameters contained in the vector $w(c)$ as control parameters of a stochastic dynamical system. So the question is about how the system responds to modifications of the control parameters which are produced exogenously. This point of view is strongly related to the problem of the interplay between self organization and central control of the urban system. Indeed all the processes changing the components of the vector state $v(c)$ are endogenous and are produced by decision processes of a set of agents (population, business, and companies). On the contrary the processes changing the vector of the control parameters $w(c)$ are related to centralized decision coming from the administration. So this distinction is also connected to the distinction between top-down processes (exogenous) and bottom-up processes (endogenous).

### 3.3. Evolution rules

In the CA theory a rule $R$ is a function associating to each cell $c$ and to each configuration $S_{U(c)}$ of its neighbourhood $U(c)$ a state $s$ belonging to the state space $\Sigma$ of the cell. The state $s = R(c, S_{U(c)}) \in \Sigma$ is the configuration of the cell at the discrete time $t + \Delta t$, $S_{U(c)}$ being the state of its neighbourhood at the time $t$.
To construct a stochastic CA one usually has to specify a set $R_1, \ldots, R_k$ of rules together with numbers $p_1, \ldots p_k \in [0,1]$, $\sum_{i=1}^{k} p_i = 1$ (these numbers are interpreted as probabilities and in general are state depending). At each step the rule $R_j$ is used with probability $p_j$. The rules of our CA differ from the usual ones because of the use of probability distributions on a continuum state space.

– **Typology of processes**

In order to define rules for our CA, we start specifying a set $A$ of processes that modify the state vector $v(c)$ of a cell $c$. For our model we have chosen the following processes as elements of $A$:

1) Occupation of free, already built, residential volume for residential use.
2) Occupation of free, already built, volume of kind I.
3) Occupation of free, already built residential volume for use as a office.
4) Abandoning of residential volume
5) Abandoning of commercial volume of kind I
6) Abandoning of offices
7) Construction of a residential building on a free terrain
8) Construction of a commercial building of kind II on a free terrain
9) Construction of a factory on a free terrain
10) Transformation of free terrain in agriculture
11) Transformation of agriculture in free terrain

In the above described specification of the model we have $A = \{1, \cdots, 11\}$. A process (event) of the kind $\alpha \in A$ will be also called an $\alpha$-process ($\alpha$-event).

– **Goods and resources of an urban process**

To specify exhaustively a transformation $\alpha \in A$ we have to give $n(\alpha)$ continuum parameters $\pi \in \mathbb{R}^{n(\alpha)}$ that describe the process $\alpha$ quantitatively. We will call these parameters the *goods* produced by the transformation $\alpha$.

We will explain this point in some details only for the process 7 (the other cases are described in an similar way). A rough description of a new residential building in the frame of our model requires the specification of its volume $\overline{V}$, of the covered surface $\overline{S}$, of the total surface $\overline{\Delta}$ associated to the building and of the internal subdivision of spaces, that is the volume $\overline{V}^A$ destined for residences or offices and the commercial volume CI: $\overline{V}^{CI} := \overline{V} - \overline{V}^A$. The surface $\overline{\Delta}$ takes into account not only the covered surface but all the space occupied by the building, including gardens, private parking and little streets for the access. It is a very important quantity because it enables to get some information about the average typology of the buildings located in the cell and it enables to evaluate correctly the amount of surface resource consumed by the building. In most cases $\overline{\Delta}$ can be identified with the surface of the parcel of terrain where the building has been built. We group together these parameters to form a vector $\pi = (\overline{V}, \overline{S}, \overline{\Delta}, \overline{V}^A) \in \mathbb{R}^{n(\alpha)}$, with $\alpha = 7$ and $n(\alpha) = 4$. The effect of an event of the kind $\alpha = 7$ on the state vector $v(c)$ of the cell $c$ can be described in the following way:

$$\begin{cases} v_1(c) = V^A \to V^A + \overline{V}^A \\ v_2(c) = V^{CI} \to V^{CI} + (\overline{V} - \overline{V}^A) \\ v_9(c) = S \to S + \overline{S} \\ v_{10}(c) = \Delta^A \to \Delta^A + \overline{\Delta} \\ v_{11}(c) = \Delta \to \Delta + \overline{\Delta} \end{cases} \qquad (3.3.1)$$

all the other components of the state vector remain unchanged. From (3.3.1) we can see that the vector $\gamma_7$, depending on $\pi = (\overline{V}, \overline{S}, \overline{\Delta}, \overline{V}^A)$, defined by:

$$\gamma_7\left(\overline{V},\overline{S},\overline{\Delta},\overline{V}^A\right):=\left(\overline{V}^A,\overline{V}-\overline{V}^A,0,...,0,\overline{S},\overline{\Delta},\overline{\Delta}\right)\in\mathbb{R}^d \qquad (3.3.2)$$

gives the variation of the state vector caused by the transformation $\alpha=7$. Similar expressions can be constructed for all other processes. In the processes 1-6 the good involved is only the volume related to the corresponding land use. In the process 8 the goods are the commercial volume of kind II, $\overline{V}^{CII}$ the covered surface $\overline{S}$ and the occupied surface $\overline{\Delta}$. In the process 9 we consider respectively the surface covered by the factory, its volume and the surface totally occupied by its surrounding infrastructures. For the processes 10 and 11 we consider the amount of surface transformed by the event.

In general we call $\gamma_\alpha\left(\pi_1,...,\pi_{n(\alpha)}\right)\in\mathbb{R}^d$, the variation of the state vector of the cell caused by the transformation $\alpha$. The complete evolution rules are obtained as superimpositions of elementary rules like (3.3.1). We will illustrate this point following the example of the variable 11, the total surface $\Delta$ occupied by the different land uses in a cell $c$. One starts considering all the processes that affect this variable. From the list of the elementary processes $A$ in section 3.3 we see that the relevant processes are those given by $\alpha\in\{7,\cdots,11\}$. The processes 7 – 10 increase the occupied surface $\Delta$ creating new buildings or new cultivated surface; the process 11 returns free surface converting agriculture and hence decreases the value of $\Delta$. This motivates the following evolution rule for the 11$^{th}$ component of the state vector

$$v_{11}(c,t+\Delta t)=v_{11}(c,t)+\sum_{\alpha=7}^{10}\gamma_\alpha\left(\pi_1^\alpha,...,\pi_{n(\alpha)}^\alpha\right)-\gamma_{11}\left(\pi_1^{11}\right) \qquad (3.3.3)$$

In general each process $\alpha\in A$ can be seen as the production of one or more goods, through the employment of other spatial extensive quantities: the *resources*. For example the creation of a new volume $\overline{V}$ causes a reduction, at the next time step $t+\Delta t$, of the volume resource $I_M^e(c)\cdot S_T-V(c,t)$ (see (2.2.1)) available in the cell in accordance with the urban plan; the same is true for good $\overline{S}$ and the surface resource $I_M^c(c)\cdot S_T-S(c,t)$. As a further example consider the occupation process 1-3; in this case the resource is represented by the available volume in the cell for the considered use.

The goods $\pi$ obviously have a stochastic nature, due to the random behaviour of the agents, whose probability distribution will be introduced in section 3.4. We will call good and resources, in general, all the spatial quantities that are respectively produced or consumed in a given process. These concepts are thus relative to a given process: a good produced in a process can be the resource employed in another process.

In mathematical language every transformation $\alpha\in A$ is described by a family $\Pi_t^i$, $i=1,...,n(\alpha)$ of time dependent random variables (the goods) with values $\pi^i$ in a time dependent interval $\left[m_t^i,M_t^i\right]\subseteq\mathbb{R}$; this interval is the set of all the allowed values for the good $\Pi_t^i$ taking into account the constraints due to the urban plan, to the availability of resources and to structural conditions as for example the minimum surface that can be covered by a certain kind of building or the maximum volume that can be built on a given surface. In the previous example $\Pi_t^3=\overline{S}$ and the resource is the interval $\left[S_0,I_M^c(c)\cdot S_T-S(c,t)\right]$, where $S_0$ is the minimum surface that can be covered by a residential building. Here, for clarity, we explicitly indicate the dependence on $(c,t)$ of the state variable $S=S(c,t)$.

### 3.4. Probability distributions associated to the automaton

In this section we introduce the probability distributions which regulate the application of the rules like (3.3.1) and hence the stochastic dynamics of the CA Let us consider a transformation $\alpha\in A$; as we sketched above, a basic idea of the model is, intuitively speaking, that the more the state of the neighbourhoods of a cell $c$ is in favour of this transformation $\alpha$, the higher are the intensities of the $\alpha$-events which will happen at the next step in the cell $c$. More precisely if the state of the neighbourhoods favours an $\alpha$-process, then we will have *a high rate of production of goods* associated to $\alpha$. But in our situation this production is mediated by the administration which has to check planning restrictions. Hence we will have a high number of requests for the transformation $\alpha$ and, depending on the available resources, only some of them will be approved and realized. From this point of view it is natural to ask that this "number of requests" is a Poisson process. In the following saying that "*an event happen in the cell*" we mean that "to the administration a request is arrived for an event concerning the cell".

## − Counting the number of events: Poisson processes

The next element that must be defined in order to complete the description of the rules for the stochastic CA is the probability density $p_t^\alpha(\pi,n)$ (with respect to Lebesgue measure $d\pi$ on $\mathbb{R}^{n(\alpha)}$) that $n$ events of the kind $\alpha \in A$ and with goods $\pi \in \mathbb{R}^{n(\alpha)}$ happen in the cell during the time interval $[t, t+\Delta t)$. In general these probability densities depend on the configuration of the system in the neighbourhoods associated to the cell and on some global variable depending on the state of the system, as we will see in the section 4 (here we have omitted this dependence to simplify the notations)

$p_{n,B}^\alpha := \int_B p_t^\alpha(\pi,n) \cdot d\pi$ give thus the probability that $n$ events of type $\alpha$ take place with goods in the (measurable) subset $B$ of the space $\mathbb{R}^{n(\alpha)}$.

The possibilities of choice for the probability densities $p_t^\alpha(\pi,n)$ can be strongly restricted analyzing more deeply the nature of the processes which happen in a given cell $c$.

Indeed, let us consider a given cell $c$ and a short time interval $[t, t+\Delta t)$. It is possible that, during this interval, nothing happens in $c$ or, on the contrary, that we have one or more events, each characterized by some discrete label $\alpha \in A$ and a set of goods $\pi \in \mathbb{R}^{n(\alpha)}$. An appropriate question about the assignment of the probability densities $p_t^\alpha(\pi,n)$ for the transitions rules is the following:

*Starting from the current state of the automaton, what is the probability $p_{n,B}^\alpha$ to have $n$ events of the kind $\alpha \in A$ and goods $\pi \in B \subseteq \mathbb{R}^{n(\alpha)}$ in the cell $c$ during the time $\Delta t$ ?*

It is reasonable to assume that the probabilities $p_{n,B}^\alpha$ are given by a Poisson distribution (we are indeed about in the same situation that we find when we want to model the process counting the number of calls received in a call-centre or the number of customers that arrive in a shop during a given time, see, e.g. (see Kingman, 1993 for basic notions on Poisson processes). Let $N_{\alpha,B}(c,t)$ be the stochastic variable that counts the number of events of the kind $\alpha$ that happen in the cell $c$ during the time interval $[t, t+\Delta t)$, with continuum parameters $\pi \in B \subseteq \mathbb{R}^{n(\alpha)}$ (in the following we will also use the simplified notation $N_{\alpha,B}(c)$ if the time dependence is clear from the context).

We recall that the main condition for the use of a Poisson distribution is the independence of the events that are counted during the time $\Delta t$. We can assume that this condition is fulfilled if $\Delta t$ is rather short, so that the following conditions are true:

1) during the time interval $[t, t+\Delta t)$ the information about the decision of the agents does not spread out in the system
2) the number of events that happen during the time $\Delta t$ is small enough so that the change of the state of the system does not significantly affect the strategy of the agents.

With these assumptions, suitably formalized, we necessarily have the following probability distributions for the counting variables:

$$P\left[N_{\alpha,B}(c,t) = n\right] = \frac{1}{n!}\exp\left(-\lambda_{\alpha,B}(c,t)\Delta t\right) \cdot \left(\lambda_{\alpha,B}(c,t)\Delta t\right)^n \tag{3.4.1}$$

where the positive-valued functions $\lambda_{\alpha,B}(c,t) \geq 0$, depending on the cell $c$ through the configurations of the system and hence on the time $t$, are the *intensities* of the Poisson process. Roughly speaking $\lambda_{\alpha,B}(c,t)$ is the velocity with which the events happen in time, indeed it is measured in 'number of events per unit time'. Note that we are using here an innocuous abuse of language with respect to the mathematical theory of Poisson processes (Kingman, 1993) where the intensity is the quantity $\lambda_{\alpha,B}(c,t) \cdot \Delta t$ and where usually $\Delta t = 1$.

The conditions stated above give us relevant, quasi-empirical criteria for the choice of $\Delta t$. Generally speaking the actual choice of $\Delta t$ depends on the size of the system, and the stated conditions are usually fulfilled if this quantity is not greater than a few days.

One can raise a natural objection: many of the processes listed above are distributed along a finite time interval which starts with the presentation to the administration of a request for a given work, and ends with the realization of the work and thus the subsequent modification of the state of the cell. This matter of fact necessarily introduces a delay time that separates the counting events, described by distributions (3.4.1), from their accomplishment, described by expressions like (3.3.1). We shall not consider delay times in this model, leaving this important improvement to subsequent works. Hence we will implicitly assume that the administration introduces the same delay time between the requests and the accordance of the permission to accomplish them. This simplification is not very strong if we think that our model includes endogenously "ordinary transformations" only, leaving the extraordinary ones to exogenous changes (see "Local parameters" in section 3.2).

In the next section we will make a further step in the construction of the intensities $\lambda_{\alpha,B}(c)$.

– **Evaluating the intensities of events**

If the time step $\Delta t$ is sufficiently small, we can suppose that all the Poisson processes (3.4.1) are independent and hence the intensity of the sum of all these counting processes is the "sum" of all the intensities (see Kingman, 1993). This is only a heuristic reasoning because we have a continuum set of processes, one for each $\pi \in \mathbb{R}^{n(\alpha)}$ and thus the previous term 'sum' is not precise, but it serves as motivation for the following definitions.

Indeed, in order to construct concretely the intensities $\lambda_{\alpha,B}(c)$ of the Poisson processes entering (3.4.1) we introduce a set of non negative, measurable functions $\lambda_\alpha$, $\alpha \in A$, depending on the cell $c$ and on the continuum parameters $\pi$ associated with the process $\alpha$:

$$\begin{aligned}\lambda_\alpha : \Gamma \times \mathbb{R}^{n(\alpha)} &\to \mathbb{R}_+ \\ (c,\pi) &\mapsto \lambda_\alpha(c,\pi) \geq 0\end{aligned} \tag{3.4.2}$$

These functions are interpreted as densities (with respect to Lebesgue measure on $\mathbb{R}^{n(\alpha)}$) for the intensities $\lambda_{\alpha,B}(c)$ so that, for any (measurable) $B \subseteq \mathbb{R}^{n(\alpha)}$:

$$\lambda_{\alpha,B}(c) := \int_B \lambda_\alpha(c,\pi) \, \mathrm{d}\pi \tag{3.4.3}$$

When $B$ coincides with the whole space $\mathbb{R}^{n(\alpha)}$ of the continuum parameters, we obtain the *counting intensity (of type $\alpha$ in $c$)* $\lambda_\alpha(c) = \lambda_{\alpha,\mathbb{R}^{n(\alpha)}}(c)$ for the events of the kind $\alpha$, no matter the value of the parameters. We can intuitively interpret (3.4.3) as the sum of a continuum set of intensities, as we said in the initial heuristic reasoning. Therefore the equation (3.4.3) together with the expression (3.4.1) enables us to interpret heuristically the quantity $\lambda_\alpha(c,\pi)\,d\pi$ as the counting intensity for events of the kind $\alpha$ and with continuum parameters $\bar\pi$ belonging to the infinitesimal $n(\alpha)$-dimensional interval $[\pi,\pi+d\pi]\subseteq \mathbb{R}^{n(\alpha)}$.

Of course we have not really solved the problem to define the intensity $\lambda_{\alpha,B}(c)$, because we have still the problem to define $\lambda_\alpha(c,\pi)$. In the following section we will see that for this aim we only need to define the global intensity $\lambda_\alpha(c)$ and the probability distribution of the goods of the transformation $\alpha$.

– **Relationships between intensities and goods**

In order to introduce probability densities for the goods produced in a $\alpha$-process let us define the following (positive) functions $\beta_\alpha(c,\pi)$:

$$\beta_\alpha(c,\pi) := \begin{cases} \dfrac{\lambda_\alpha(c,\pi)}{\lambda_\alpha(c)} & \text{if } \lambda_\alpha(c) > 0 \\ 0 & \text{if } \lambda_\alpha(c) = 0 \end{cases} \qquad (3.4.4)$$

We then have from (3.4.4) $\int_{\mathbb{R}^{n(\alpha)}} \beta_\alpha(c,\pi)\cdot d\pi = 1$ if $\lambda_\alpha(c)\neq 0$, where we have used (3.4.3) for $B=\mathbb{R}^{(\alpha)}$ and the relation $\lambda_\alpha(c) := \lambda_{\alpha,\mathbb{R}^{n(\alpha)}}(c)$ discussed in the previous subsection. It is easy to show (Vancheri et al. a, in preparation) that if we consider the $n(\alpha)$-dimensional infinitesimal interval $[\pi,\pi+d\pi]\subseteq \mathbb{R}^{n(\alpha)}$, then the quantity $\beta_\alpha(c,\pi)\cdot d\pi := \beta_\alpha(c,\pi_1,...,\pi_{n(\alpha)})\cdot d\pi_1\cdot ...\cdot d\pi_{n(\alpha)}$ gives the probability to have goods $(\bar\pi_1,...,\bar\pi_{n(\alpha)})\in[\pi,\pi+d\pi]$ for an $\alpha$-event that has taken place in the cell $c$. E.g. for $\alpha=7$, $\beta_7(c,V,S,\Delta,V_A)\cdot dV\cdot dS\cdot d\Delta\cdot dV_A$ gives the probability that there is a request for the construction of a residential building with volume $\bar V\in[V,V+dV)$, covered surface $\bar S\in[S,S+dS)$, total surface associated to the building $\bar\Delta\in[\Delta,\Delta+d\Delta)$ and residential volume $\bar V^A\in[V^A,V^A+dV^A)$.

In the case $\lambda_\alpha(c)=0$ the previous normalisation condition on $\beta_\alpha(c,\pi)$ is not fulfilled, but in this case no event can happen in the cell and thus we do not need to define any probability distribution $\beta_\alpha(c,\pi)$ for the goods.

Thus the problem to define $\lambda_\alpha(c,\pi)$ is traced back to the definition of the global intensity $\lambda_\alpha(c)$ and of the probability density $\beta_\alpha(c,\pi)$. The first one is discussed in section 4, whereas the second one is discussed in the next section.

– **Probability distributions for the goods of an urban transformation**

Generally speaking, the choice of the distributions $\beta_\alpha(c,\pi)$ can be formalized as follow: we firstly define the resources (time depending intervals in $\mathbb{R}$) of the transformation $\alpha$ (see also the discussion at the end of the section 3.3:

$$\begin{aligned}
R_\alpha^1(t) &:= \left[m_\alpha^1(t), M_\alpha^1(t)\right] \\
R_\alpha^2(\pi_1,t) &:= \left[m_\alpha^2(\pi_1,t), M_\alpha^2(\pi_1,t)\right] \\
&\dots \\
R_\alpha^{n(\alpha)}(\pi_1,...,\pi_{n(\alpha)-1},t) &:= \left[m_\alpha^{n(\alpha)}(\pi_1,...,\pi_{n(\alpha)-1},t), M_\alpha^{n(\alpha)}(\pi_1,...,\pi_{n(\alpha)-1},t)\right]
\end{aligned} \qquad (3.4.5)$$

Herein $m_\alpha^i, M_\alpha^i : D_\alpha^i \to \mathbb{R}$ are functions giving the minimum respectively the maximum value of the $i$-th resource and whose domains are:

$$D_\alpha^i := \left\{ (\pi_1, ..., \pi_{i-1}, t) \in \mathbb{R}^i \,\middle|\, \pi_1 \in R_\alpha^1(t), \pi_2 \in R_\alpha^2(\pi_1, t), ..., \pi_{i-1} \in R_\alpha^{i-1}(\pi_1, ..., \pi_{i-2}, t) \right\} \quad (3.4.6)$$

Finally we choose probability densities $\beta_\alpha^i(-; c^i, d^i) : \mathbb{R} \to \mathbb{R}$, $i = 1, ..., n(\alpha)$, where with $[c^i, d^i] \subseteq \mathbb{R}$ we indicate in general the support of the probability densities:

i. $\beta_\alpha \left( c, \pi_1, ..., \pi_{n(\alpha)} \right) := \prod_{i=1}^{n(\alpha)} \beta_\alpha^i \left( \pi_i; m_\alpha^i(\pi_1, ..., \pi_{i-1}, t), M_\alpha^i(\pi_1, ..., \pi_{i-1}, t) \right)$ \quad (3.4.7)

ii. $\beta_\alpha^i \left( \bullet; m_\alpha^i(\pi_1, ..., \pi_{i-1}, t), M_\alpha^i(\pi_1, ..., \pi_{i-1}, t) \right)$ is the conditional probability density of the good $\Pi^i$ given the values $\Pi^k = \pi_k$, $k = 1, ..., i-1$, $i = 1, ..., n(\alpha)$.

Note that these definitions depend on the sequence of goods $\Pi^1, ..., \Pi^{n(\alpha)}$ used by the agents "to realize" the $\alpha$-transformations.

For a practical implementation of the model we chose the densities $\beta_\alpha^i(\bullet; c^i, d^i)$ from the family of beta distributions or of positive polynomials on the interval $[c^i, d^i]$ (the latter is computationally better than the former). A non trivial problem is represented by the way we follow to evaluate the parameters of the chosen distributions $\beta_\alpha^i(\bullet; c^i, d^i)$. If the chosen distributions depends on two parameters (this is the case, e.g. of the beta distributions) we can evaluate them through their relation with the mean and the variance of the distribution. The mean and the variance can depend on the trend of the agents to standardize new buildings to previously existing ones or by the tendency to fully use the available resources in accordance the rules of the urban plan. Anyway in the construction of the model one has to expect to have sufficiently rich local parameters (as discussed in section 3.2) to evaluate the parameters distribution of $\beta_\alpha^i(\bullet; c^i, d^i)$ in a meaningful way.

## 4. Constructing counting intensities.

After having described the general mathematical and conceptual setting of our model, we will outline to some extent the general structure of the intensities for the process and some general concepts introduced in order to describe the interactions among the different land uses.

### 4.1. From counting of events to rate of production of goods.

In section 3.4 we have decomposed the intensity $\lambda_\alpha(c, \pi)$ for an $\alpha$-process occurring in a cell $c$ as the product of a counting intensity $\lambda_\alpha(c)$ and a probability density $\beta_\alpha(c, \pi)$ for the production of the goods associated to $\alpha$ (equation (3.4.4)). In the same section we have discussed the probability density $\beta_\alpha$. Now we will sketch the general approach followed to relate the counting intensities $\lambda_\alpha(c)$ with quantities of interest in connection with applications.

− **Mean rate of production of a good**

We start by remarking that, when a concrete model has to be built, the function $\lambda_\alpha(c)$ is not the most convenient quantity to start with. The mean number of residential buildings produced in a given time, for example, does not contain any direct information about the residential volume that has been effectively produced: a little house and a huge residential building are considered indeed both as a unit in the counting. So, for example, it is much more relevant to consider the mean rate of production of residential volume. This quantity can be indeed directly connected with the demand on new residential spaces due to the demographical pressure acting on the city. So it seems to be more meaningful to give criteria for the mean rate of production of a good in a given process $\alpha$ rather than for the intensity $\lambda_\alpha(c)$ of the counting of $\alpha$-events.

Bearing these consideration in mind, let us consider a cell $c$ and a process $\alpha$ with the related counting intensity $\lambda_\alpha(c)$. Let us consider further an extensive, non negative, quantity $\xi_\alpha$ produced in the $\alpha$-process (that is an extensive function of the goods $\pi \in \mathbb{R}^{n(\alpha)}$). For $\alpha = 7$ we can consider for example $\xi_\alpha := \pi_1 = \overline{V}$, the volume produced in the $\alpha$-event.

The intensity of the counting $\lambda_\alpha(c)$ and the mean rate $\Lambda_\alpha(c)$ of production of the good $\xi_\alpha$ in $\alpha$-processes can be connected through an elementary property of Poisson processes: the expected value of the number of events counted during a small time $\Delta t > 0$ is given by $\lambda \cdot \Delta t$, where $\lambda$ is the intensity of the process. Let us consider an infinitesimal volume $d\pi$ centred in $\pi \in \mathbb{R}^{n(\alpha)}$. The expected number of $\alpha$-events during the time step $\Delta t$ with continuum parameters in $[\pi, \pi + d\pi)$ is given by $\lambda_\alpha(c,\pi) \cdot d\pi \cdot \Delta t$. If we multiply this quantity by the amount $\xi_\alpha(\pi)$ of the considered good produced in the event we obtain the mean amount of the good $\xi^\alpha$ produced in $c$ during $\Delta t$ for events with continuum parameters in $d\pi$. Integrating further on $\mathbb{R}^{n(\alpha)}$ and dividing by $\Delta t$ we obtain the rate $\Lambda_\alpha(c)$:

$$\Lambda_\alpha(c) = \int_{\mathbb{R}^{n(\alpha)}} \lambda_\alpha(c,\pi) \xi_\alpha(\pi) d\pi \tag{4.1.1}$$

Substituting $\lambda_\alpha(c,\pi)$ in the (4.1.1) through (3.4.4) we obtain:

$$\Lambda_\alpha(c) = \lambda_\alpha(c) \cdot \overline{\xi}_\alpha(c) \tag{4.1.2}$$

where $\overline{\xi}_\alpha(c) = \int_{\mathbb{R}^{n(\alpha)}} \beta_\alpha(c,\pi) \xi_\alpha(\pi) d\pi$ is the expected value of the amount of good $\xi_\alpha$ produced in a $\alpha$-process in the cell $c$.

The first step in constructing a concrete model thus consists in the definition for each $\alpha$-process of the mean rate of production $\Lambda_\alpha(c)$ for a suitable chosen good $\xi_\alpha$. These rates have to be chosen as functions of the state of the system in the neighbourhoods of the cell $c$ and of some global variables depending on the state of the system (see equation (4.1.4) ). After having chosen a probability density $\beta_\alpha(c,\pi)$ for the goods (see section 3.4), we can obtain the quantities $\overline{\xi}_\alpha(c)$ and hence the counting intensities $\lambda_\alpha(c)$ from (4.1.2):

$$\lambda_\alpha(c) = \frac{\Lambda_\alpha(c)}{\overline{\xi}_\alpha(c)} \tag{4.1.3}$$

It is easy to see that we have $\overline{\xi}_\alpha(c) = 0$ only for $\beta_\alpha(c,\pi) = 0$, $\forall \pi \in \mathbb{R}^{n(\alpha)}$. But in this case we have $\lambda_\alpha(c) = 0$ (equation (3.4.4) ), and it is impossible to have $\alpha$-events in $c$.

− **Global contribution to the intensities .**

The rate of production $\Lambda_\alpha(c)$ of a good $\xi_\alpha$ depends in general on the configuration of the system in a suitable set of neighbourhoods of the cell $c$. Besides local, neighbourhood dependent contributions, we have to take into account also effects acting on the system at a global level. This fact has been pointed out explicitly by White and Engelen, who fix exogenously (or quasi-exogenously) the total growth of the number of cells of a given kind during a step, leaving to the local rules to decide where these new cells will be located. Indeed it is not reasonable to assume, for example, that the rate of growth of the population or of the commercial surfaces be only given by the local conditions found in the neighbourhoods of the cells. Global features of the system, as e.g. its size and the global demand on a certain kind of space, seem to be more relevant than local ones to determine the global rate of growth of a given land use. The local conditions are more relevant in order to decide the spatial distributions of the land uses created during the growth process. In fact the co operation and the antagonism among land uses related with spatial proximity and the correspondent level of rents act as local forces to determine the location of the land uses.

We will follow this general idea about the separation of global and local effects, with the important difference that the global rates of growth of the land uses will not be introduced separately as a quasi exogenous sector, but will be introduced directly in the expressions for the counting intensities $\lambda_\alpha(c)$. The advantages of this approach will be the possibility to have a stronger connection between local and global aspects in the decision processes of the agents.
In order to accomplish this task we write the rate of production of the good relevant for a process $\alpha$ in the following way:

$$\Lambda_\alpha(c) = \Lambda_\alpha^G(e) \cdot p_\alpha(c) \qquad (4.1.4)$$

where $\Lambda_\alpha^G$, the global term, is a function of the global configuration $e \in E$ of the system (see (3.2.2)), and $p_\alpha(c)$ is a local term depending on the cell $c \in \Gamma$ through the state of its neighbourhood and fulfilling the following two conditions: i) $p_\alpha(c) \in [0,1] \; \forall c \in \Gamma$ and ii) $\sum_{c \in \Gamma} p_\alpha(c) = 1$.

The global rate of growth of the good $\xi_\alpha$ associated to the $\alpha$-process on the whole system is obtained summing up the contributions (4.1.4) for each cell. We obtain then that the global rate of growth is given by the term $\Lambda_\alpha^G(e)$. In this way the local term $p_\alpha(c)$ represents the fraction of the global intensity $\Lambda_\alpha^G(e)$ which is attributed to the cell $c$. In order to construct local terms $p_\alpha(c)$ that fulfil the properties i) and ii) stated above, we define for each cell $c \in \Gamma$ and for each process $\alpha \in A$ an $\alpha$-potential $P_\alpha(c) \geq 0$, which describes the degree of attractiveness of the cell $c$ with respect to the process $\alpha$. We obtain the local terms just normalizing the potentials: $p_\alpha(c) := P_\alpha(c)/T_\alpha(e)$, where $T_\alpha(e) = \sum_{c \in \Gamma} P_\alpha(c)$ is the normalisation factor.

The global rate of growth $\Lambda_\alpha^G(e)$ and the local terms $p_\alpha(c)$ play in a certain sense the same role respectively of the global exogenous system of control of the dynamics and of the local rules based on the potentials in the automaton of White and Engelen.
Substituting (4.1.4) into (4.1.3) and using (3.4.4) we obtain for the densities $\lambda_\alpha(c,\pi)$ of the counting intensities $\lambda_\alpha(c)$:

$$\lambda_\alpha(c,\pi) = \Lambda_\alpha^G(e) \cdot p_\alpha(c) \cdot \frac{1}{\overline{\xi_\alpha}(c)} \cdot \beta_\alpha(c,\pi) \qquad (4.1.5)$$

The general expressions (4.1.5) for the counting intensities contain two extreme cases: a) the dynamics without any global term, where only the neighbourhood relations among cells are relevant and b) the completely exogenous control of the global rates.
The case a) is obtained when $\Lambda_\alpha^G(e) = \nu_\alpha \cdot T_\alpha(e)$, where $\nu_\alpha$ are positive constants. In this case one obtains immediately from (4.1.5) $\lambda_\alpha(c,\pi) = \nu_\alpha \cdot P_\alpha(c) \cdot \beta_\alpha(c,\pi) / \overline{\xi_\alpha}(c)$ and thus only local contributions, proportional to the potentials, appear in the intensities. This assumption has some not realistic consequences, as for example the fact that a sudden change in the urban plan making a subset of cells unavailable for a given $\alpha$ process has no direct effect on the rates in the remaining cells. We would expect a partial redistribution of the intensities on the cells that remain active.

The case b) is obtained when $\Lambda_\alpha^G(e) := f_\alpha(t)$, where $f_\alpha$ is a given function of the time $t$ that does not depend on the configuration $e$ of the system. It is not difficult to see that, putting $f_\alpha(t) = \gamma_\alpha^1 \cdot \exp(\gamma_\alpha^2 \cdot t)$, where $\gamma_\alpha^1$ and $\gamma_\alpha^2$ are positive constants, we obtain a global rate of growth $\Lambda_\alpha^G(e)$ of $\xi_\alpha$ which is proportional to the size of the system in the $\xi_\alpha$-sector. This assumption corresponds to a CA with fully constrained global dynamics.

But the most interesting cases correspond to a not trivial dependence of the global terms $\Lambda_\alpha^G(e)$ on a suitable chosen set $J_k(e)$, $k = 1, \cdots, M$, of global observables with a clear economical, urban and demographic meaning. If we assume, for example, that the rate of growth of an urban system can be represented as a function $f(n)$ of its population $n$, we can define for the process $\alpha = 1$: $\Lambda_1^G(e) = f_1(J_1(e))$, where $J_1(e) = V_{u,\text{tot}}^R(e)/a$ is a rough evaluation of the population of the city through the total occupied residential volume $V_{u,\text{tot}}^R(e) = \sum_{c \in Z} V_u^R(c)$ and the estimated volume per inhabitant $a$.

A more refined definition of $\Lambda_1^G(e)$ can be obtained introducing a larger set of global observables describing other aspects of the urban system concerning for example the quality of life and the demand on new occupation in the industrial sector and in services.

As a further example let us consider the process $\alpha = 8$, construction of commercial spaces of kind II. It is meaningful, at least as a first approximation, to assume that the rate of growth of commerce II is connected to a global demand on commerce in the city. This demand can be roughly described through an index $J_2(e) = V_{u,\text{tot}}^R(e) - \eta \cdot \left( V_{u,\text{tot}}^{CI}(e) + V_{u,\text{tot}}^{CII}(e) \right)$ where the first term is the estimated total population of the city and the second one is the global amount of commercial volume. The exogenous parameter $\eta > 0$ is defined in such a way that values $J_2(e) > 0$ are associated to configurations with a global demand on commercial surfaces and values $J_2(e) < 0$ with configurations with a global excess in commerce respect to the population. Putting $\Lambda_8^G(e) = \gamma_1 \cdot \exp(\gamma_2 \cdot J_2(e))$ we have that the total growth of the commerce II is linked to the growth of the population. This link can be made more or less strong by varying the value of the positive parameter $\gamma_2$.

## 4.2. Constructing the potentials with fuzzy logic

The most important quantity in the construction of the rates of counting are the potentials $P_\alpha(c)$ that describe the degree of attractiveness of a cell $c$ for $\alpha$-processes. It is well known that the level of the rent is one of the most important forces that determine the location of the land uses on the urban space. The rent is connected to the attractiveness of a given urban site for a land use, a factor that depends strongly on the local conditions in the part of the city surrounding immediately the terrain. Thus the location of the land uses depends both on forces that tend to concentrate the demand of terrain where it is especially convenient to run a given kind of activity and on forces connected with the level of the rent that act to spread out the spatial distribution of the activities. The strong dependence of attractiveness and rent on local conditions make the problem suitable for an approach through Cellular Automata.

We measure the attractiveness of a terrain for a given land use by means of a semi quantitative method (we will call it, translating from German, "positional method") elaborated by Nägely and Wengen (Nägely and Wengen, 1997) to evaluate the values of buildings and the related rents and prices. The basic assumption is that the value of a building located in a given part of the land can be conventionally separated into two contributions: the value of the building itself, that can be identified with the building costs, and the value of the terrain, which depends on the relation of the terrain itself with the land surrounding it and with the rest of the land.

Nägely and Wengen construct a check list that takes into account many relevant factors for the evaluation of the value of a terrain, as the centrality of the area where the terrain is located, the relation with the transport network, the accessibility to services, green parks and cultural activities and the distribution of land uses around the terrain. For each land use X a score $Q^X$ between 0 and 10 is assigned to the terrain where the building is or will be located, according to the qualitative and semi-quantitative criteria carefully indicated in the check list through an exhaustive list of examples. The ratio of the value of the building to the value of the terrain is obtained by means of an empirical positive valued function $q$ of the score $Q^X$. Let $W_T$ and $W_I$ be the values attributed respectively to the terrain and to the building. The total value $W$ is given then by:

$$W = W_T + W_I = q(Q^X) \cdot W_I + W_I = W_I \cdot (1 + q(Q^X)) \tag{4.2.1}$$

The rent is approximately estimated as a certain fraction $\gamma > 0$ of the total value $W$. If we indicate by $V$ the volume of the building and by $w_I$ the value of the building per unit of volume, then the rent per unit of volume will be:

$$\begin{aligned} r^X(Q^X) &= \frac{W}{V}\gamma = \frac{W_I \cdot (1 + q(Q^X))}{V}\gamma = \\ &= \frac{V \cdot w_I \cdot (1 + q(Q^X))}{V}\gamma = w_I \cdot (1 + q(Q^X)) \cdot \gamma \end{aligned} \tag{4.2.2}$$

Relation (4.2.2) enables only a rather rough evaluation of the rents because it does not take into account the effect of several factors describing the dynamics of the market in the $X$-sector. Nevertheless some basic aspect of the dynamics of the rents, for example their dependence on the relation of a terrain with the urban system, is captured. This idealized model can be considered as a first approximation of a more complete model, containing explicitly a more detailed economical sector. The potential $P_\alpha(c)$ for an $\alpha$-process is evaluated through the scores $Q^X$ and the estimated rents $r^X(Q^X)$, where $X$ runs over the land uses relevant for the process $\alpha$. We will sketch some general ideas about the method we use to accomplish this task in the last part of this section. A more detailed description of the method will be presented in Vancheri et al. b (in preparation).

In using the positional method we are faced with the problem of translating the qualitative criteria furnished by the check list, expressed in verbal terms, in quantitative operations applicable to the state vectors and to the vectors of the local parameters. This task has been accomplished using a method based on fuzzy logic and fuzzy set theory (see, e.g. Dubois and Prade, 1980; Bandemer and Gottwald, 1995).

The score associated through the positional method to a terrain located in a cell $c$ for a use $X$ can be interpreted, if normalized between 0 and 1, as a fuzzy truth value of a proposition $p^X :=$ "*The cell is attractive with respect to the use $X$*". The truth value $\mu^X(c)$ of such a proposition with respect to a cell $c \in \Gamma$ can take a continuum set of values between 0, corresponding to "absolutely false", and 1, corresponding to "absolutely true". In this way a fuzzy set $A^X$, characterized by a membership function $\mu^X$ is constructed on the universe $\Gamma$ (the cellular space). The fuzzy set $A^X$ can be meant as the set of all the attractive cells with respect to the land use $X$.

The statement $p^X$ can be seen as a complex linguistic construction made of more elementary statements $p_j^X$ about some relevant features of the neighbourhood of the cell, as indicated in the check list of the positional method. Fuzzy logic is used to analyse the structure of the main statement $p^X$, to assign truth values $\mu_j^X$ to its elementary components $p_j^X$ and to reconstruct through sentential connectives like "and", "or", "not" and linguistic modifiers like "very", "more or less" the membership function $\mu^X$ of the fuzzy set $A^X$. The score attributed by means of the positional method are given by $Q^X(c) := 10 \cdot \mu^X(c)$.

Processes $\alpha \in A$ are considered, following fuzzy decision theory, as decision making processes where one has to attain goals without violating constraints. In the most simple case $\alpha = 1$ (occupation of residential volume) the goal consists in maximizing the attractiveness of the cell where the residential volume will be occupied and the constraints are given by the necessity to pay a not to high rent and to find free residential volume in the cell Following standard ideas in fuzzy decision making one has to consider the proposition $p_1 := p^R \wedge (\neg r^R) \wedge u^R$, that can be read as *"the cell is attractive as a residential site* and *the rent is not too expansive* and *there is free residential volume available in the cell"*. The membership function $\mu^R$ associated to the proposition $p^R$ is obtained through the positional method, as mentioned above. The membership functions associated to $r^R$ (*"the rent is too expansive"*) and $u^R$ (*" there is free residential volume in the cell"*) depend respectively on the residential rent $r^R(Q^R)$ obtained through the positional method. and on the amount of not occupied residential volume. These membership functions have the shape shown in the figure 1 and 2:

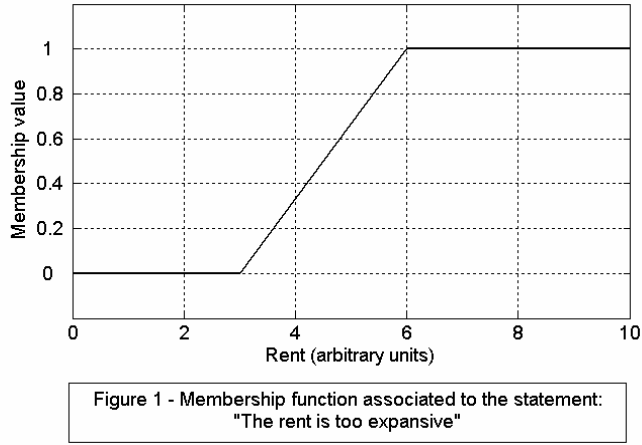

Figure 1 - Membership function associated to the statement: "The rent is too expansive"

Through fuzzy complement and fuzzy conjunction, the latter being provided by some standard $t$-norm (see Bandemer and Gottwald, 1995) the membership functions are combined to give the membership function $\mu_1$ for the statement $p_1$. We could use directly as potential the membership functions $\mu_\alpha$ obtained for each $\alpha$-process following ideas similar to those sketched above. But we prefer to define potentials through exponentiation of the membership functions:

$$P_\alpha(c) := (\mu_\alpha(c))^{\rho_\alpha} \qquad (4.2.3)$$

where $\rho_\alpha > 1$ are real exponents that enhance the attractiveness of the cells which have larger values of the membership functions $\mu_\alpha(c)$. This choice is justified by observing that the membership functions $\mu_\alpha(c)$ and the potential $P_\alpha(c)$ express respectively the evaluations of the agents about the attractiveness of a cell and the probability that the cell will be chosen as target of the decision. Small differences in the evaluation can result in strong polarisation on the most favourable case. The exponentiation has exactly the effect to increase the polarisation of the choices of the agents on the most favourable cells.

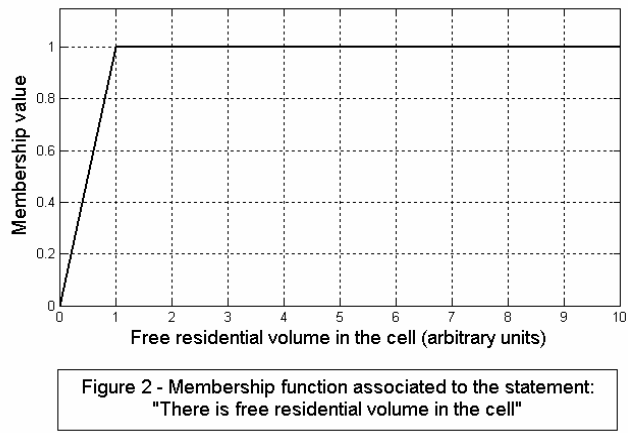

Figure 2 - Membership function associated to the statement: "There is free residential volume in the cell"

The construction by means of the positional method of the membership function $\mu^X$ for the different land uses follows ideas similar to those explained above, but it is rather involved. We will explain here only some ideas by means of examples concerning the residential use $R$. The elementary statements involved in the construction of $\mu^R$ all refer to quantities the can be expressed as very simple functions of the state vector, as for instance the mean height of the buildings (3.2.3), the density of built volume (3.2.4) the covering density (3.2.5) and many others. The elementary statement in the third example is for instance $p_1^R :=$ "*the urban space around the cell is densely covered*". For sake of simplicity we will follow in some details only this latter example. In order to take into account contributions from the cells surrounding the given cell $c$, we have to modify formula (3.2.5) in the following way:

$$I^S(c) = \frac{\sum_{c' \in \Gamma} z_{cc'}^S \cdot S(c')}{\sum_{c' \in \Gamma} z_{cc'}^S \cdot S_T(c')} \qquad (4.2.4)$$

where $S(c)$ and $S_T(c)$ are respectively the surface covered by buildings and the total surface of the cell without considering fixed land uses. The contributions are summed up with weights $z_{cc'}^S \geq 0$ that take into account the effect of distance. In order to understand better this latter point let us rewrite the expression in the following way:

$$I^S(c) = \frac{\sum_{c' \in \Gamma} z_{cc'}^S \cdot S_T(c') \cdot \frac{S(c')}{S_T(c')}}{\sum_{c' \in \Gamma} z_{cc'}^S \cdot S_T(c')} \qquad (4.2.5)$$

The covering density $I^S(c)$ associated to the cell $c$ is thus a weighted mean of the covering densities $S(c')/S_T(c')$ with weights $S_T(c') \cdot z_{cc'}^S$ depending on the distance between $c$ and $c'$ through the parameters $z_{cc'}^S$ and on the size of the cell $c'$ through the quantity $S_T(c')$. Obviously only a few number of cells $c' \in \Gamma$, the ones belonging to a neighbourhood of $c$, will contribute to the sums with coefficients $z_{cc'}^S > 0$. We will consider this aspect at the end of this subsection, when we will define the neighbourhoods more precisely.

To the elementary statement $p_1^R$ "*the urban space around the cell is densely covered*" is associated a membership function $\mu_1^R$ as shown in the figure 3.

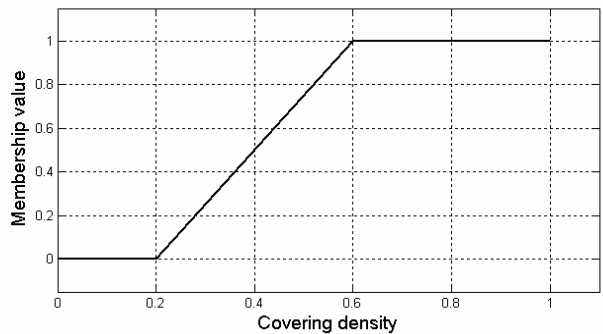

Figure 3 - Membership function associated to the statement: "The urban space around the cell is densely covered"

The second example is given by the integration of a cell in the system induced by the transport network. Let us consider for each cell $c \in \Gamma$ the nearest vertex $v_c$ and its distance $r_c$ from $c$ (see section 2.3). Let us consider further the proposition $p_2^R := p_1^C \wedge p_2^C$ that we read as *"the nearest vertex is well connected* and *the nearest vertex is not too far away from the cell"* (here the superscript C means "connection"). Membership functions $\mu_1^C$ and $\mu_2^C$ depending respectively on the integration index $i(v_c)$ (see equation (2.3.2)) and on the distance $r_c$ are defined in a natural way and hence compounded via a suitable chosen t-norm (for example the product) to give the membership function $\mu_2^R$ associated to the proposition $p_2^R := $ *"the cell is well integrated in the system"*. Using modifiers as exponentiation on $\mu_2^R$ it is possible to introduce membership function for variables as "very connected" or "rather connected" which are suited to treat the effect of connection on different land uses.

Through standard operations in fuzzy logic (t-norm, t-co norm, generalized averaging, modifiers) on membership functions like $\mu_1^R$ and $\mu_2^R$ it is possible to construct the fuzzy set $A^R$ constituted by the cells $c \in \Gamma$ which are attractive for the residential use R and the related membership function $\mu^R$.

The main advantage of this approach is given by its closeness with the criteria followed by agents in the usual processes of decision making. Another advantage is that the membership functions used to construct the potentials contain a consistent part of the parameters used in the model. The values of these parameters are not directly measurable quantities; nevertheless they reflect judgments of peoples about isolated and simple features of the space around the cells and can be chosen in a meaningful way without calibrating them through computer simulations (there are several method at disposal based on judgments expressed by experts (see Yager R.R., 1977 for an example). The number of parameters that must to be calibrated becomes in this way not too large.

At this stage we have to spend some words about the method we use to define the neighbourhoods of the cells. The weights $z_{cc'}^j$ used to define the elementary feature $j$, as for instance the covering density, are parameters to be calibrated. In order to avoid a huge number of parameters we proceed subdividing all the cells surrounding $c$ into few groups (typically three or four) and associating to the cells of each group the same weight: we obtain for each elementary feature $j$ a set of $n_j$ "rings" $U_k^j(c)$ $k=1,2,\cdots,n_j$ surrounding $c$ and characterized by a weight $z(U_k^j) > 0$. More precisely we have that $c',c'' \in U_k^j(c) \Rightarrow z_{cc'}^j = z_{cc''}^j = z(U_k^j)$. Obviously in order to obtain local interactions it is necessary to consider only a few number of cells having a weight different from zero.

In order to define neighbourhoods $U_k^j = U_k^j(c)$ in a meaningful way a criterion is needed in order to decide whether two cells contribute almost the same extent to a given interaction with the cell. The best case is when we have at disposal a meaningful distance between cells (not necessarily given by the metric distance on the urban space). In most cases the distance has to be measured through an accessibility time with respect to the relevant transport network (as already pointed out, the subway could be a good mean to reach the office but not to reach a shopping centre; in the latter case it is more convenient to use a car). Having such a measure $d$ of the distance at disposal we can define a neighbourhood of a cell $c$ as the set of all the cells located between two determined distances $d_k^j, \delta_k^j$ from $c$

$$U_k^j(c) := \{c' \in Z: \quad d_k^j \leq d_j(c,c') < \delta_k^j\} \qquad (4.2.6).$$

Let us notice that this way to construct neighbourhoods can lead in general to neighbourhoods which are spread out on the space and not regular in shape. In the first computer simulation (to be presented in Vancheri at al. b, in preparation) we have used only the metric distance to build neighbourhoods and we have employed systematically four ring of cells surrounding the central cell, the first being given by the cell itself.

# 5. Description of the algorithm

In the preceding sections we have discuss all elements which defines our model: its geometrical (i.e. cellular) and temporal structure (3.1); the state space of the cells (3.2); the evolution rules (3.3) and the probability distributions associated with the automaton (3.4). In particular we have seen how we can motivate the introduction of Poisson processes with intensities $\lambda_{\alpha,B}(c,t)$ and how these intensities can be defined using $\lambda_\alpha(c,\pi)$ (see (3.4.3) ). The subsequent problem to define $\lambda_\alpha(c,\pi)$ is traced back to the definition of the global intensity $\lambda_\alpha(c)$ and of the probability density $\beta_\alpha(c,\pi)$ through the relation (3.4.4). Finally we motivated possible assumptions on $\beta_\alpha(c,\pi)$ (see (3.4.7)) and on $\lambda_\alpha(c)$ (see section 4).

If we proceed with the most natural synchronous evolution algorithm for this CA, we have firstly to evaluate the intensities of each process in each cell. Using a Poisson distributed random numbers generator we have to extract the number of $\alpha$-events which happen in each cell. Then using a $\beta_\alpha(c,\pi)$-distributed random numbers generator we extract the goods for each transformation and finally we apply the evolution rules.

The following *asynchronous* evolution algorithm, based on the connection of the dynamics of our CA with the theory of Markov jump processes (MJP), is more efficient because it uses less Poisson extractions than the synchronous one:

1. Let us consider the CA at the beginning of a generic step $n$. Let the time be given by $t_n = (n-1)\cdot \Delta t$, where, as before, $\Delta t$ is the discrete time step. All the counting variables $N_\alpha(c,t_n) := N_{\alpha,\mathbb{R}^{n(\alpha)}}(c,t_n)$ ("number of $\alpha$-transformations which happen in $c$ during the time interval $[t_n, t_n + \Delta t)$") for $\alpha \in A$ and $c \in \Gamma$ are supposed to be independent, so that the quantity

$$N(t_n) := \sum_{\substack{\alpha \in A \\ c \in \Gamma}} N_\alpha(c,t_n) \tag{5.0.1}$$

("number of transformation of some type in some cell") is again Poisson distributed with intensity

$$\lambda(t_n) := \sum_{\substack{\alpha \in A \\ c \in \Gamma}} \lambda_\alpha(c,t_n) \tag{5.0.2}$$

We can thus answer the question *"when do the next transformation of the CA (of some type $\alpha$ in some cell $c$) happen?"* extracting an exponentially distributed random number $\theta \geq 0$ with intensity $\lambda(t_n)$. We interpret a situation in which $\theta > \Delta t$ as "the first event does not happen during the $n$-th step but in $k \cdot \Delta t \leq \theta < (k+1)\cdot \Delta t$, $k \in \mathbb{N}$, $k \geq 1$" (note that in this case the intensities do not change up to the $k$-th step because the state of the automaton does not change).

2. *"Where do this first transformation happen?"* We have proved that we must answer this question using the discrete distribution

$$\left(\frac{\lambda_c}{\lambda}\right)_{c \in \Gamma}, \text{ where } \lambda_c := \sum_{\alpha \in A} \lambda_\alpha(c) \tag{5.0.3}$$

At the end of this step of the algorithm we have a time $\theta$ and a cell $c$.

3. *"Of what type $\alpha \in A$ is the transformation that happen in the cell $c$ at the time $\theta$?"* We have proved that we must answer this question using the following discrete distribution:

$$\left(\frac{\lambda_\alpha(c)}{\lambda_c}\right)_{\alpha \in A} \tag{5.0.4}$$

At the end of this step we have a time $\theta$, a cell $c \in Z$ and a process $\alpha \in A$.

4. *"How much of the goods does the $\alpha$-transformation produce?"*. We generate the values $\pi_1,...,\pi_{n(\alpha)}$ of the goods using the distribution $\beta_\alpha(c,\pi)$ (see section 3.4).

5. At this step of the algorithm we have a time $\theta$, a cell $c$, the type of process $\alpha \in A$ and the values of the goods $\pi_1,...,\pi_{n(\alpha)}$. Using this information we calculate the new value of the time with the relation $t \to t + \theta$ and the new values of the resources available in the cell $c$. We repeat the steps 1-4 of the algorithm while the condition $t \leq t_n + \Delta t$ is fulfilled. The state of the CA, the counting intensities $\lambda_\alpha(c)$ and the probability densities $\beta_\alpha(c,\pi)$ of the goods are modified only when the whole time step is passed ($t > t_n + \Delta t$). The state of the system is changed applying the rules (see section 3.3) to each transformation happened during the time step $[t_n, t_n + \Delta t)$.

We refer to Vancheri et al. a, (in preparation) for a rigorous proof of the statements contained in the description of the algorithm. Let us notice that the algorithm sketched above is based on the assumption that there is no delay between the starting point of a process and its accomplishment (see section 3.4). This enable us to model the time evolution of the CA as in a (MJP), with exponentially distributed sojourn time on the states (see Vancheri et al. a, in preparation). The explicit introduction of a delay time is among the improvements of the model planned in the future.

## 6. Ordinary differential equations

In this section we write an ordinary differential equation (ODE) for the time evolution of the mean values of the dynamical variables. The rigorous derivation of this (approximated) ODE involves the use of a Markov semigroup associated to the stochastic evolution of the system and the related Kolmogorov forward equation. Non markovian effects connected with the history of the system and delay times connected with the time duration of the processes (see section 3.4) can hardly be managed in a MJP based approach. We have faced these problems using random differential equations (RDE) (based on the use of the mean forward derivative by Nelson) instead of ordinary ones (see Giordano et al., in preparation). We have two main advantages in the use of RDE instead of ODE. The first one is the possibility to use differential equation under very general conditions (e.g. non Markov stochastic evolution or randomly distributed delay times). The second one is connected with the possibility to derive from RDE the probability distributions for the dynamical variables (in particular of expected values and variances). In this section we will limit ourselves in writing down an ODE and justifying it intuitively.

Let us start considering a cell $c \in \Gamma$ and the related state vector $v(c,t)$ at a given time $t$. The probability distributions (3.4.1) can be expanded in a Taylor series in the time step $\Delta t$:

$$\begin{cases} B = [\pi, d\pi) \\ P[N_{\alpha,B}(c,t) = 0] = 1 - \lambda_{\alpha,B}(c,t) \cdot \Delta t + o(\Delta t) \\ P[N_{\alpha,B}(c,t) = 1] = \lambda_{\alpha,B}(c,t) \cdot \Delta t + o(\Delta t) \\ P[N_{\alpha,B}(c,t) \geq 2] = o(\Delta t) \end{cases} \quad (6.0.1)$$

From (6.0.1) we see that we only need to consider the cases $N_{\alpha,B}(c,t) = 0,1$ when the time step $\Delta t$ is very short. Let us consider now the system in a given state $e \in E$ at the time $t$. The variation of the state vector after a jump of type $\alpha \in A$ and goods $\pi \in R^{n(\alpha)}$ (we will indicate shortly this jump by $(\alpha,\pi)$) has been introduced in the section 3.3 and indicated with the symbol $\gamma_\alpha(\pi)$ (see (3.3.3) and the related discussion). The probability density for a $(\alpha,\pi)$-jump is approximately given by:

$$dp_\alpha(c,\pi) = \lambda_\alpha(c,\pi) \cdot d\pi \cdot \Delta t \qquad (6.0.2)$$

where we have applied (6.0.1) to a small subset $[\pi, \pi + d\pi) \subset \mathbb{R}^{n(\alpha)}$ and we have discarded all the contributions from more than one jump. The mean variation $\overline{\Delta v}(c,t)$ of the state vector during the time step $\Delta t$ can intuitively be obtained summing up all the possible jump amplitudes $\gamma_\alpha(\pi)$ each multiplied through the related probability $dp_\alpha(c,\pi)$. This leads to the equality:

$$\overline{\Delta v}(c,t) = \Delta t \cdot \sum_{\alpha \in A} \int_{R^{n(\alpha)}} \lambda_\alpha(c,\pi) \cdot \gamma_\alpha(\pi) \cdot d\pi \qquad (6.0.3)$$

Equation (6.0.3) can be used to evaluate the expected configuration of the cell $c$ at the time $t + \Delta t$ *under the assumption that the configuration at the time $t$ is $v(c,t)$*. This latter assumption prevents to use formula (6.0.3) recursively in time in order to evaluate the time evolution of the expected configuration $\overline{v}(c,t)$ of the cell. The expected value of the state vector after the second step of the recursion should be evaluated taking into account as starting points for the (6.0.3) all the possible positions reached by the state vector after the first step. This makes the formula for the evaluation of $\overline{v}(c,t)$ tremendously complicated after few jumps. A strong simplification can be obtained starting at each step from the expected position reached in the previous step. This enable to use (6.0.3) recursively to obtain a ODE in the limit for $\Delta t \to 0$:

$$\frac{d\overline{v}(c,t)}{dt} = \lim_{\Delta t \to 0} \frac{\overline{\Delta v}(c,t)}{\Delta t} = \sum_{\alpha \in A} \int_{R^{n(\alpha)}} \lambda_\alpha(c,\pi) \cdot \gamma_\alpha(\pi) \cdot d\pi \qquad (6.0.4)$$

The assumption used to write an ODE (6.0.4) is intuitively justified only if the stochastic paths covered by the state vector during its time evolution remain with a high probability close to the most probable path $\overline{e}(t)$. A more rigorous treatment of the mean value equation shows explicitly that (6.0.4) holds only when this condition (opportunely formulated) is fulfilled (Vancheri et al. a, in preparation; Giordano et al., in preparation).

## 7. Conclusions

In this paper we have introduced a new model for modelling urban systems based on a continuum valued stochastic CA. We have focussed especially on the general modelling strategy followed for this kind of CA, leaving to subsequent papers both the applications to real case studies and the theoretical development of the model. The main steps followed in modelling urban systems with our type of CA can be summarized in the following way:

- The space state of a cell is constituted by a set of extensive quantities as built volumes and occupied surfaces for different land uses (section 3.2 ).
- The evolution rules are given by the probability distribution of stochastic, Poisson distributed jumps corresponding to several kinds of processes on the urban space. Each jump is labelled by a set of continuum extensive parameters, called goods, that describes quantitatively the process (section 3.4 ).
- The typology of each cell is reconstructed using fuzzy logic methods through the state vectors in a suitable set of neighbourhoods (section 4 ).

- The dynamics of the system is investigated either using direct simulation of stochastic paths using the algorithm described in section 5 or solving the ordinary differential equation as described in section 6).

The main point we deal with in the next steps of our research will concern the application of the model to real case studies. This involves especially the problem of parameter calibration and data acquisition and management. The first problem will be faced using both comparison of simulated and empirical configurations by means a meaningful metric on the configuration space and fuzzy decision theory techniques.

Furthermore we shall improve our model introducing dynamical variable describing several kinds of flows on the urban space (e.g. workers, customers directed from residences to shopping centres) in order to take into account long range interactions taking place in the urban space. This task will be accomplished introducing a higher level cellular decomposition of the urban space, based on macro-cells obtained grouping together a certain number of ordinary cells.

The third direction of development is more theoretical and concerns the use of random differential equations to investigate the stochastic evolution of the CA. Ordinary differential equations can be employed indeed only under rather restrictive conditions, when the stochastic fluctuations around the most probable path can be disregarded. On the contrary random differential equations enable to explore the fully stochastic dynamics of the system and to search e.g. for phase transitions induced by noise.

## 8. Acknowledgements

We thank Arch. Mirko Galli for having suggested us the use of the positional method in order to define the potentials associated to the processes and for the subsequent useful and stimulating discussions. His help has been precious also in the realisation of the artificial urban environment used for the computer simulations that we will present in Vancheri et al. b (in preparation).